# HARMONI at ELT:
# Reoptimization of the NGSS natural guide star sensors system for MCAO compatibility

Dohlen K.[1], Le Mignant D.[1], Bonnefoi A.[1], Dubbeldam M.[2], Floriot J.[1], Funes Vecino I.[3], Jocou L.[4], Richaud Y.[1], Carlotti A.[4], Challita Z.[1], Costille A.[1], Dunn A.[2], King D.[2], Malone D.[2], Mercant Rubio G.[3], Morris T.[2], Sauvage J.F.[5,1], Staykov L.[2], Viera Curbelo T.A.[6]

1: Aix Marseille Univ, CNRS, CNES, LAM, Marseille, France; 2: Durham University, Centre for Advanced Instrumentation, United Kingdom; 3: Centro de Astrobiología (CAB), CSIC-INTA, Spain; 4: Université Grenoble Alpes, CNRS, IPAG, France; 5: DOTA, ONERA, 13330 Salon de Provence, France; 6: Instituto de Astrofísica de Canarias (IAC), Universidad de La Laguna, Spain

## ABSTRACT

HARMONI is the first-light, adaptive optics assisted, near-IR integral field spectrograph for the ELT. It covers a spectral range from below 800nm to 2400nm with resolving powers from 3000 to 7000 and spatial sampling of 25mas and 6mas. It can operate in three adaptive optics (AO) modes: single conjugate AO (SCAO), high-contrast AO (HCAO), and multi-conjugate AO (MCAO). The project is resuming its final design phase after a rescope design phase that has lasted two years. This paper describes the impact of the rescope on the natural guide star sensors (NGSS) system, imposing wide-ranging changes in particular to the low-order wavefront sensors (LOWFS) previously optimized for LTAO operation as well as on the external support and enclosure (ESE) whose external interfaces have changed. The high-contrast module (HCM) is also deeply affected due to the modified plate scale of the instrument, and the single-conjugate AO sensor (SCAOS) benefits from a reduced patrol field requirement. Description of the changes will be given together with the associated system analysis involving requirements flow-down, revisited functional analysis, and updated product breakdown structure. Elements of the new system management structure and work breakdown will also be included.

## 1. INTRODUCTION

HARMONI [1],[2] is a near-infrared integral field spectrograph providing one of the ELT's core spectroscopic capabilities. It uses image slicers to provide spectra for each one of its 204x152 spaxels covering a single contiguous field. To benefit fully from the sensitivity and spatial resolution offered by the nearly 40 m diameter pupil of this giant telescope, HARMONI will be able to work at the diffraction limited with more that 90% sky coverage thanks to laser-assisted adaptive optics. It also offers a high-contrast mode aiming at early exploitation of the capacities of this giant telescope to reach closer and deeper that current facilities on 10-meter class telescopes.

Following difficulties of convergence towards the final design review, a decision was taken to proceed to a re-evaluation of science goals [3] leading to a rescope of the instrument. In addition to reducing the number of instrument settings in terms of spatial scales and spectral resolution options, it was decided to move the instrument to the second port of the MORFEO to benefit from its multi-conjugate adaptive optics (MCAO) functionality in replacement of the laser-tomographic adaptive optics (LTAO) earlier designed into the instrument. In addition to improved sky-coverage thanks to the improved AO correction of multiple guide stars, this change allowed removal of one of the HARMONI systems, the laser guide-star sensors (LGSS) and the focal-plane relay system (FPRS). The instrument now consists of three opto-mechanical systems, the integral field spectrograph (IFS), a large cryostat mounted on a rotating table, the natural guide-star sensors (NGSS), a thermally stabilized container mounted on top of the IFS, and the calibration and relay system (CARS) providing the large mirror folding the MORFEO beam vertically, calibration equipment, and the structure carrying these elements above the IFS and NGSS.

The NGSS [4] has maintained most of its system architecture through the rescope, still consisting of three wavefront sensor subsystems, LOWFS, SCAOS, and HCM, as well as the internal support bench, ISB, and the external support and enclosure, ESE. After a brief description of the rescoped HARMONI instrument and the position of NGSS within it, this paper recalls the functionalities of NGSS and each of its subsystem and details the changes and optimizations effectuated.

## 2. RESCOPED HARMONI AND ITS CONTEXT

HARMONI is now located at the second output port of MORFEO, opposite the MICADO instrument located at its first output port, see Figure 1. MORFEO projects the reimaged telescope image plane onto a flat surface 14 m downstream of its exit pupil and with a sufficient back-focal distance to allow the beam to be folded downwards by the HARMONI-provided M12 fold mirror, pass through the NGSS and into the IFS much in the same way as the earlier HARMONI configuration. While the changes in optical interfaces (pupil location and image curvature) lead to minor re-optimisation of the optical designs of wavefront sensors and IFS pre-optics, the passage of the beam from the MORFEO output port, sloping upwards to M12 caused somewhat painful mechanical design implementations to avoid vignetting.

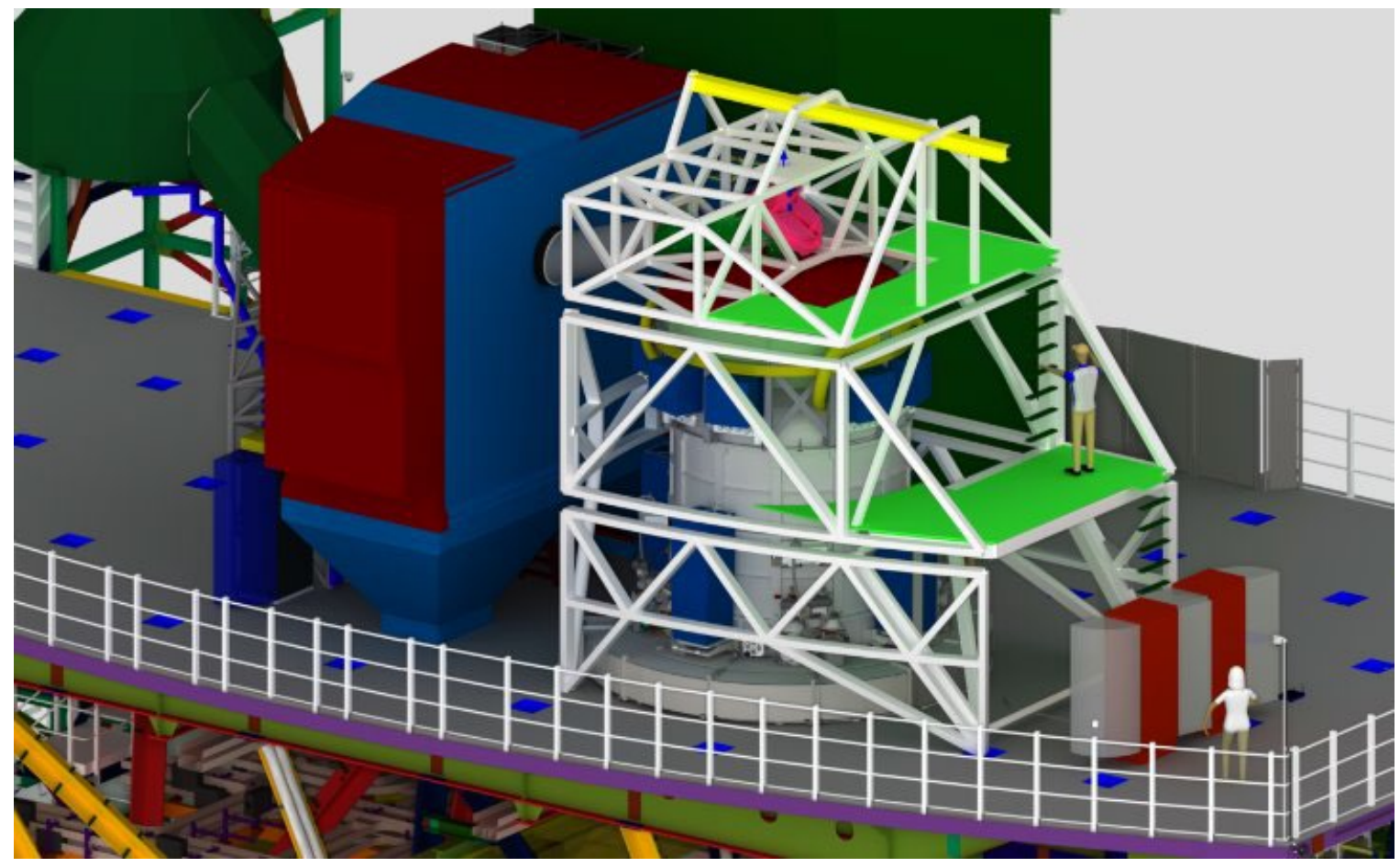

Figure 1. Conceptual view of rescoped HARMONI on the Nasmyth platform at the second output port of MORFEO.

Instead of four spatial spaxel scales, 4x4 mas, 10x10 mas, 20x20 mas, and 30x60mas, the rescoped HARMONI only has two scales, 6x6 mas and 25x25 mas, abandoning Shannon sampling of the Airy disk in the H band in favour of improved signal-to-noise ratio. Judging Shannon sampling of the residual speckle pattern essential in high-contrast mode, additional scale change is provided by the NGSS HCM module as described below.

## 3. RESCOPED NGSS

### 3.1 Entrance window location

One of the most critical requirements of NGSS is to allow absolute pointing of the science field with respect to known guide stars to milli-arcsecond (mas) precision and to maintain stability during exposures. This requires extreme-precision actuators for object selection, but also a highly stable thermal environment. In the original design, the four mirrors of the relay system were required to be cold to minimize thermal background, and to avoid multiplying optical windows, a cold and thermally stable environment was shared between the two systems. Rescoped HARMONI has abandoned the weight put on thermal background since MORFEO mirrors are not cooled, leading to a range of different architectural options, see Figure 2.

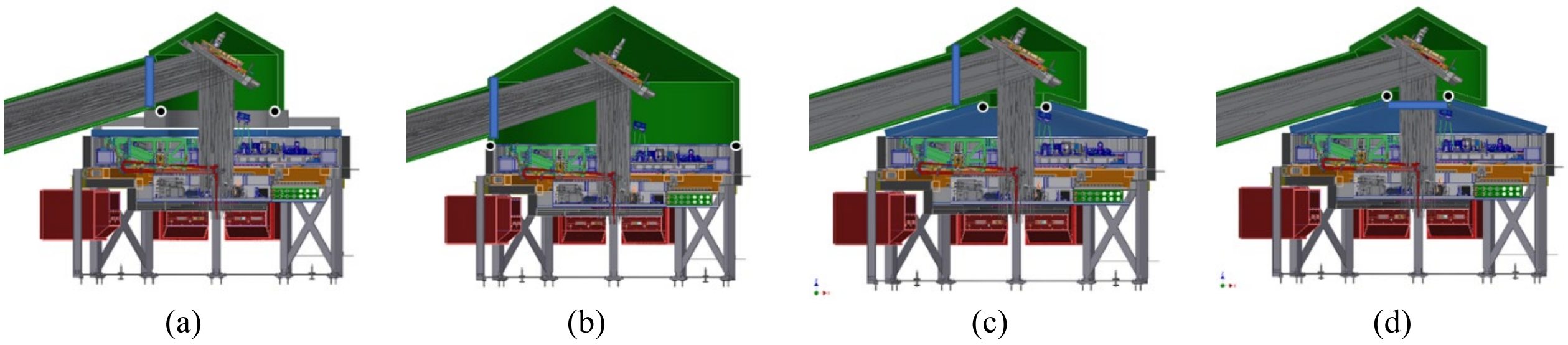

Figure 2. Some of the architectural options considered for the rotational joint between NGSS and the fixed M12 volume. (a) The M12 is carried by NGSS on a large bearing allowing opposite rotation to that of the cryostat so it stays fixed with respect to MORFEO. The volume is shared, tightness being ensured by a rotary joint. (b) The M12, carried by an external support structure, shares a large closed volume with the NGSS. (c) Same as (b) but minimizing the M12 volume and the diameter of the rotary joint. (d) Same as (c) but with a window mounted into NGSS separating the volumes.

Options (a) and (b) were soon discarded due to complexity and the high friction of the large rotary seal. Options (c) and (d) were further considered as part of a family of options concerning window locations, see Figure 3. A crude trade-off matrix allowed choosing between these options, taking into account and noting from -1 to 1, five important factors:

- NGSS stability, representing the capacity to maintain a thermally stable environment,
- M12 cooling, representing the issues involved with a shared volume
- Dust accumulation, on the window itself and on other surfaces.
- Window cleaning access
- Number of windows, affecting transmission

Based on these criteria, the optimal location of a single window is found to be W3, at the entrance of NGSS. Use of two windows score equally high, but in view of the cost and complexity of manufacturing and installing such windows, as well as the obvious cost in transmission, these have been dismissed. From a system design point of view, the W3 option is also clearly optimal, allowing for a permanent and clear separation between two separate systems, manufactured and delivered by two different consortium partners. Indeed, the shared volume present in the old design, although judged manageable, was definitely a challenge, in particular in terms of guaranteeing near-hermetic tightness of the rotating seal. In the current design, the seal will still be expected to provide tightness for dust ingress, but constraints in terms of thermal and humidity ingress are relaxed. The choice does imply a shared volume with MORFEO, however. Discussions with the MORFEO team is ongoing to determine the implications of this, in terms of requirements on thermal insulation and cleanliness.

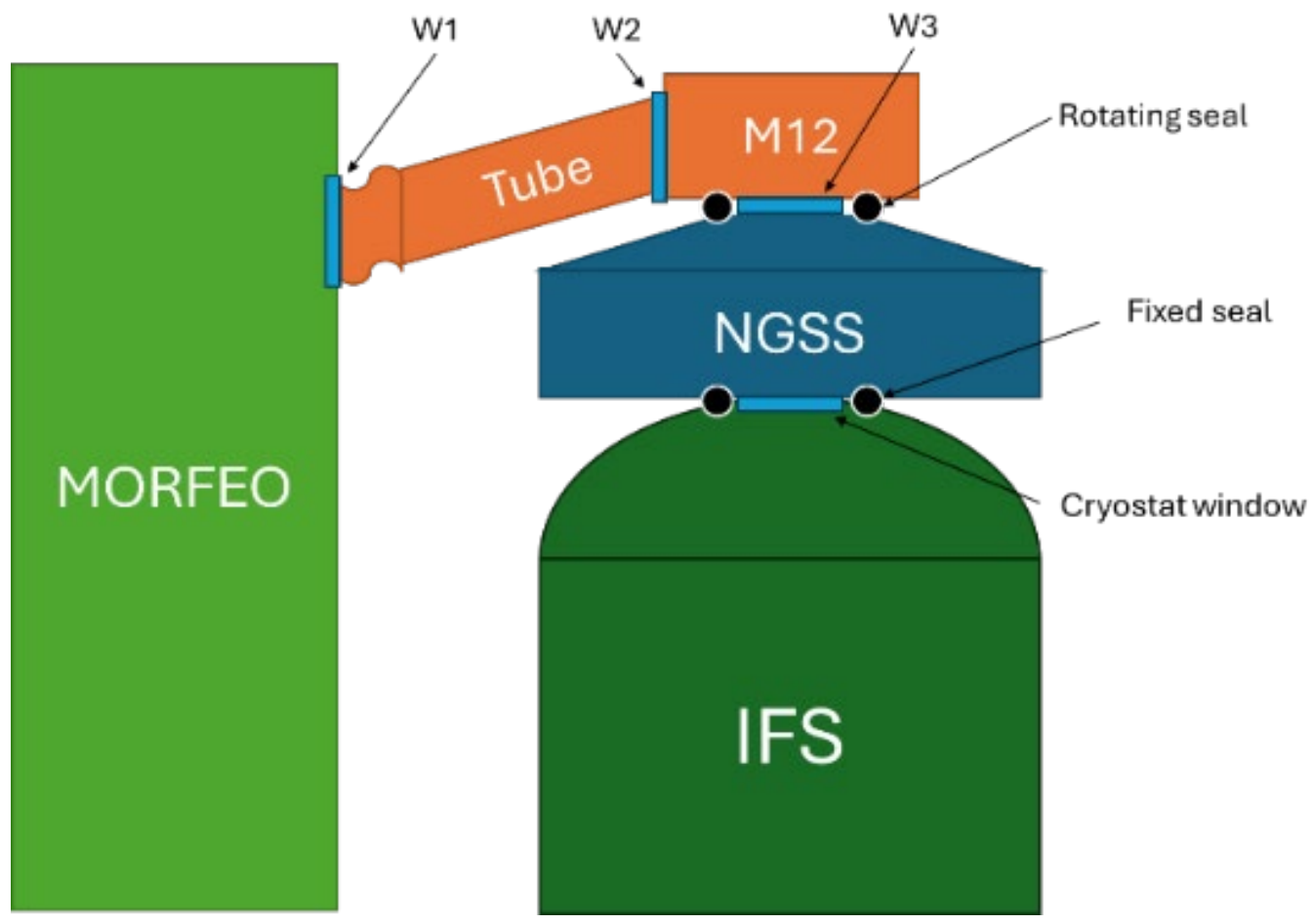

Figure 3. Schematic illustration of NGSS window location options: W1: at the interface with MORFEO, W2: at the entrance of the M12 volume, W3: at the entrance of NGSS.

Table 1. Trade-off table for location of the NGSS window.

| | No window | W1 | W2 | W3 | W1 + W2 | W2 + W3 | W1 +W3 | W1 + W2 + W3 |
|---|---|---|---|---|---|---|---|---|
| **NGSS stability** | -1 | 0 | 0 | 1 | 0 | 1 | 1 | 1 |
| **M12 cooling** | -1 | 0 | 0 | 1 | 0 | 1 | 1 | 1 |
| **Dust accumulation** | -1 | 0 | 0 | 0 | 0 | 1 | 1 | 1 |
| **Window cleaning** | -1 | -1 | -1 | 0 | -1 | 0 | 0 | 0 |
| **Number of windows** | 1 | 1 | 1 | 1 | 0 | 0 | 0 | -1 |
| **Sum** | **-3** | **0** | **0** | **3** | **-1** | **3** | **3** | **2** |

### 3.1.1 Dust accumulation

One of the most critical issues of the selected window location is its tendency to accumulate dust. It is well known that horizontal upwards looking surfaces are significantly more prone to dust accumulation, with a generally assumed ten times higher fallout rate than for vertical surfaces. Dust fallout for a horizontal surface in different cleanroom environments is classified in [5], see Figure 4. We have previously studied the cleanliness achieved within the NGSS volume, where critical surfaces like the cryostat entrance window, also upwards looking, will be difficult if not impossible to reach for regular cleaning operations. The NGSS volume is therefore designed to be near-hermetically sealed, ensuring a slight overpressure by injecting cleaned and dried air. The air cleaning system uses compressed air supplied by the telescope which is passed through a membrane type filter [6]. An experiment was set up at ATC to estimate the cleanliness achieved using this equipment, see Figure 5. The experiment indicates that a class 5 environment can be reached, for which a fallout rate of 0.1% aerial coverage per year is expected. For the NGSS window, exposed to the M12 volume which is connected to the MORFEO volume via a large-diameter tube, ensuring such over pressure would be impossible, so we will have to live with a shared volume. Does that mean we should expect a cleanliness level equivalent to that of the ELT, specified to "ISO 7.2", which according to Figure 4 produces a fallout rate of ~3% aerial coverage per year? Although the top surface of the NGSS window will be made accessible for regular cleaning operations, it is clear that minimizing the fallout rate is seen as highly desirable. It is expected that the location of the M12 volume at the sealed end of a long, upwards-sloping tube should limit dust ingress considerably, both for large particles, for which gravitational fallout dominates, and for the smallest particles, for which Brownian motion dominates. Avoiding upwards draft will be important, but since pressure should be quite efficiently equalized thanks to the large opening into MORFEO, creation of such a draft is also unlikely. We are therefore confident that fallout rates could be considerably lower than 3%/year.

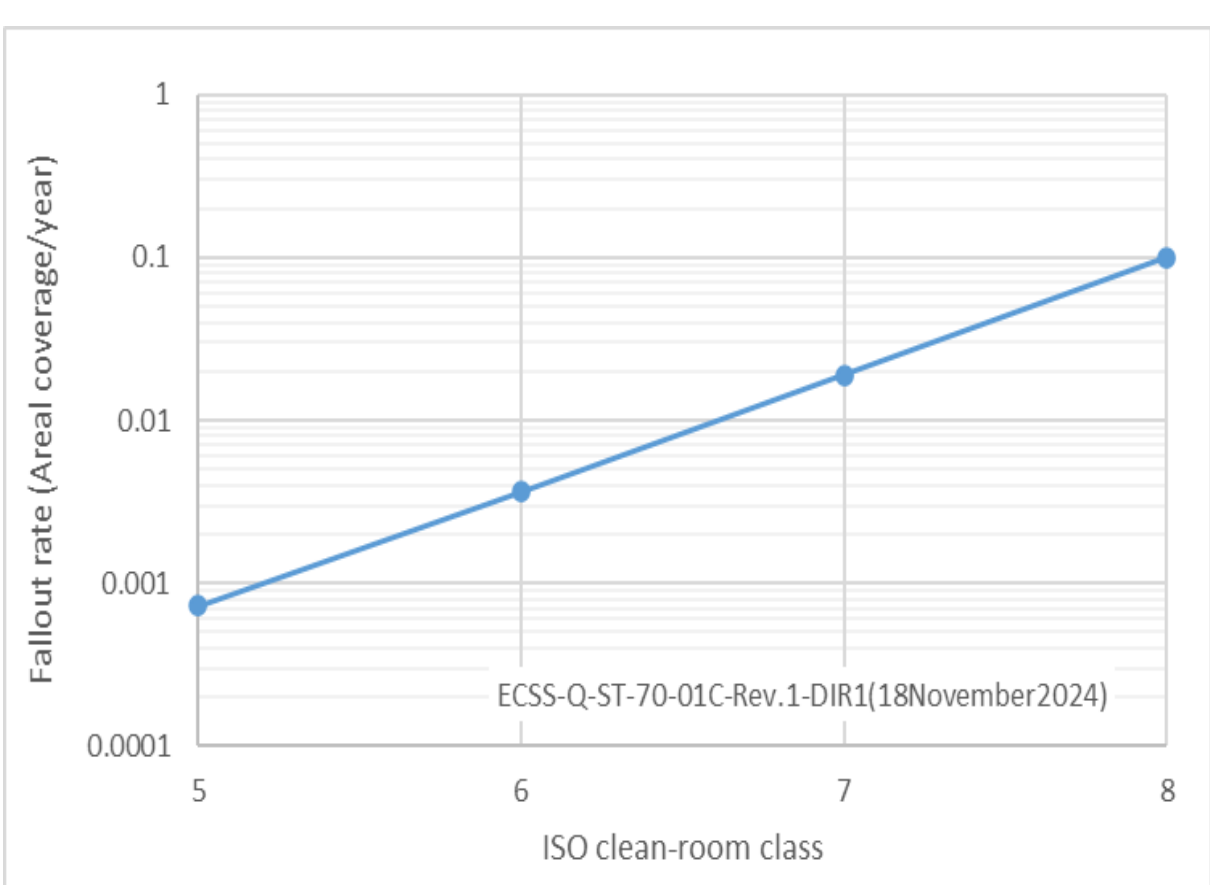

Figure 4. Fallout rate in terms of fractional aerial coverage per year of exposure to a given ISO clean-room class for an upwards-looking surface [5].

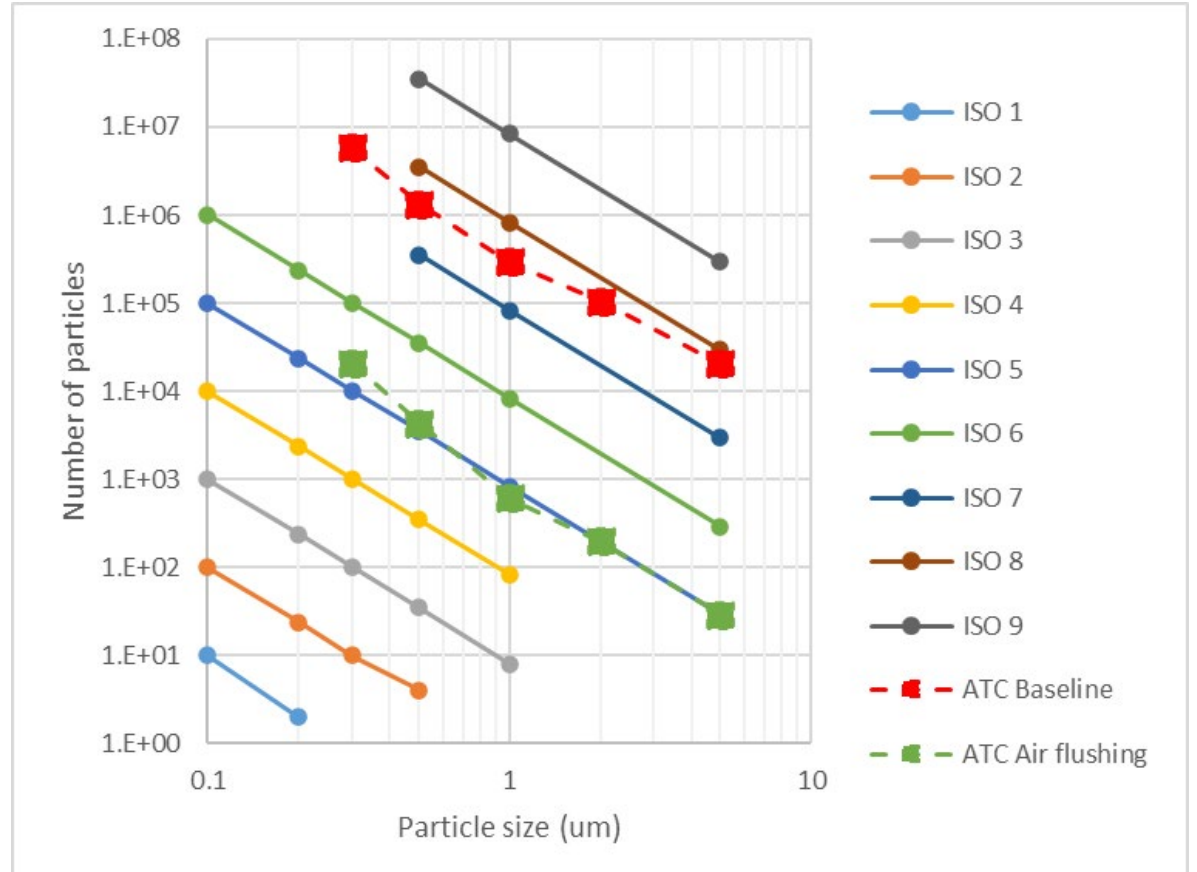

Figure 5. Particle counts obtained in the ATC air cleaning experiment without (red broken line) and with (green broken line) the air filtering equipment installed, compared with ISO clean room standard particle counts (solid lines).

### 3.2 Window temperature: fogging or turbulence

In the earlier HARMONI design, the NGSS temperature was set to minimize thermal background. It was originally set to -15 degC, but increased to +2 degC to avoid difficulties due to frosting, ice accumulation within cracks in the cover etc. This increase also made qualification of mechanisms and devices easier and it relaxed constraints on design and chiller power. In the new HARMONI, the background argument falls, and so it would be interesting to increase the temperature further, close to the median observatory temperature, assumed to be around 7 degC, to further reduce average power consumption. As an added advantage, this would lead to a warmer window, less prone to fogging. Still, to maintain the very stringent stability requirements for NGSS wavefront sensors, required to maintain milli-arcsecond pointing precision to reach its world coordinate system (WCS) requirements, it has been judged necessary to maintain the temperature fixed rather than floating, leading to the window being warmer than ambient half the time, hence leading to local turbulence.

A crude model for window external temperature has been set up, ignoring radial heat transfer, ie representing the case of an infinitely large window, see Figure 6. In this case, heat transfer to the ambient on either side of the window through radiation and convection is balanced by the heat conducted through the window, allowing calculation of inner and outer window temperature. It is difficult to precisely determine the convection coefficient in this case, since both volumes are closed with minimal air movement. Also, we wish to operate in turbulence-free conditions. We therefore consider zero convection as a conservative case leading to the largest differential surface temperatures. In this condition, Figure 7 plots external window temperature as function of dome temperature for the cases of 2 and 7 degC internal temperature. It also plots dew temperature at 40, 60, and 80% relative humidity (RH), knowing that the ELT operational up to RH=80%. Above that level, the dome is expected to be closed and heated to avoid risk of condensation. It appears that a risk of condensation occurs for dome temperatures down to ~9 degC if we maintain NGSS at 2 degC, while there is only a marginal risk if we up the temperature to 7 degC.

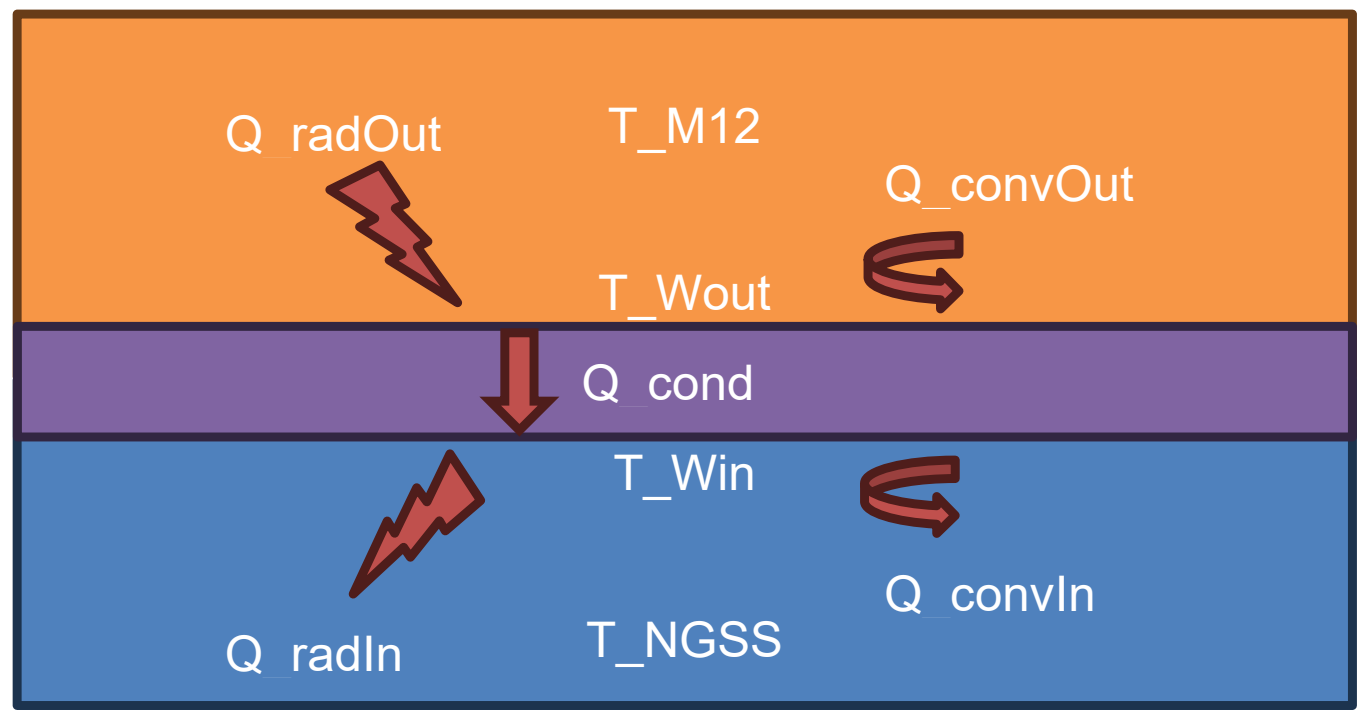

Figure 6. NGSS window thermal model.

A hot plate creates buoyancy-driven upwards flow of air which tends to create turbulence when the plate is sufficiently hot. The transition between laminar and turbulent flow is characterized by the Rayleigh number [7]. In the absence of wind, the Rayleigh number is given by:

$$R = \frac{g\beta\Delta T L^3}{\nu\alpha},$$

where g=9.81 m/s$^2$ the gravitational constant, β=0.0037 K$^{-1}$ the coefficient of thermal expansion of air, ν=1.5x10$^{-5}$ m$^2$/s the cinematic viscosity, α=2.2x10$^{-5}$ m$^2$/s the thermic diffusivity, L thickness of the air layer. Plugging in the constants, the Rayleigh number is

$$R = 1.1\text{x}10^8\ \Delta T\ L^3.$$

While R=1.7x10$^3$ is considered to mark the passage into a laminar regime, R>10$^6$ is considered a turbulent regime. Figure 9 shows examples of turbulence at different Rayleigh numbers [8].

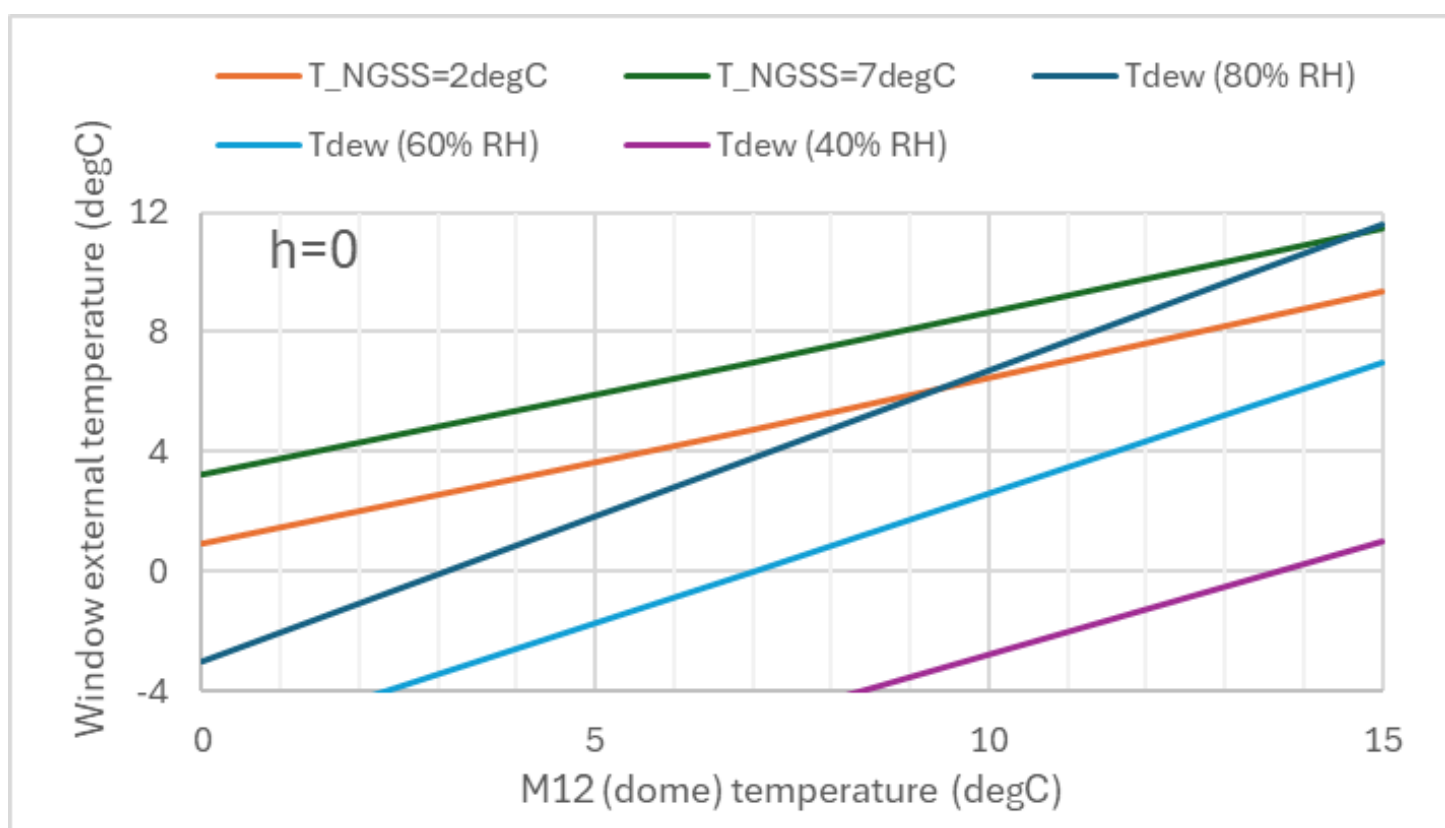

Figure 7. Window external temperature as a function of M12 temperature for the case of 2degC (orange) and 7degC (green) internal NGSS temperature for the case of zero convective exchange. The blue line shows dew point temperature in the case of RH=80%.

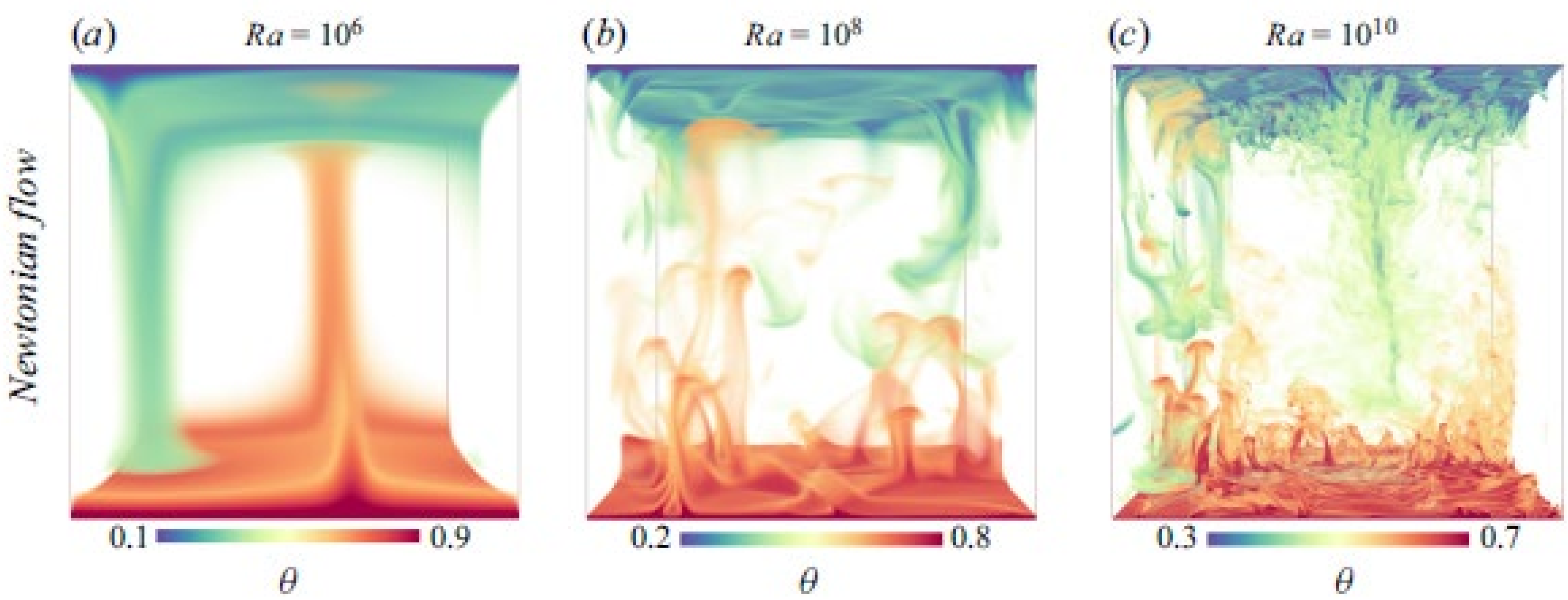

Figure 8. Instantaneous temperature fields of Newtonian flows for different Rayleigh numbers, *R*. Reproduced from [8].

Positive temperature difference between window surface and M12 will occur whenever ambient temperature falls below that of the NGSS internal, see Figure 9. In the case where NGSS is stabilized to 7degC (orange curve), the difference reaches +3degC.

Assuming the median height between the NGSS window and the M12 mirror L = 0.5 m and for a worst case of $\Delta T = +3K$, the Rayleigh number is 4.1e7, clearly within the range of turbulent flow. The Rayleigh number is plotted against ambient temperature for operating temperatures of 2 and 7 degC in Figure 10, indicating that turbulence is likely to occur very close to operating temperature.

Unfortunately, therefore, there is no ideal internal temperature for NGSS. While maintaining the current 2 degC baseline temperature will give rise to temperatures on the outer window surface below the dew point in hot and humid weather (T>9degC, RH>70%) increasing the temperature will create local turbulence.

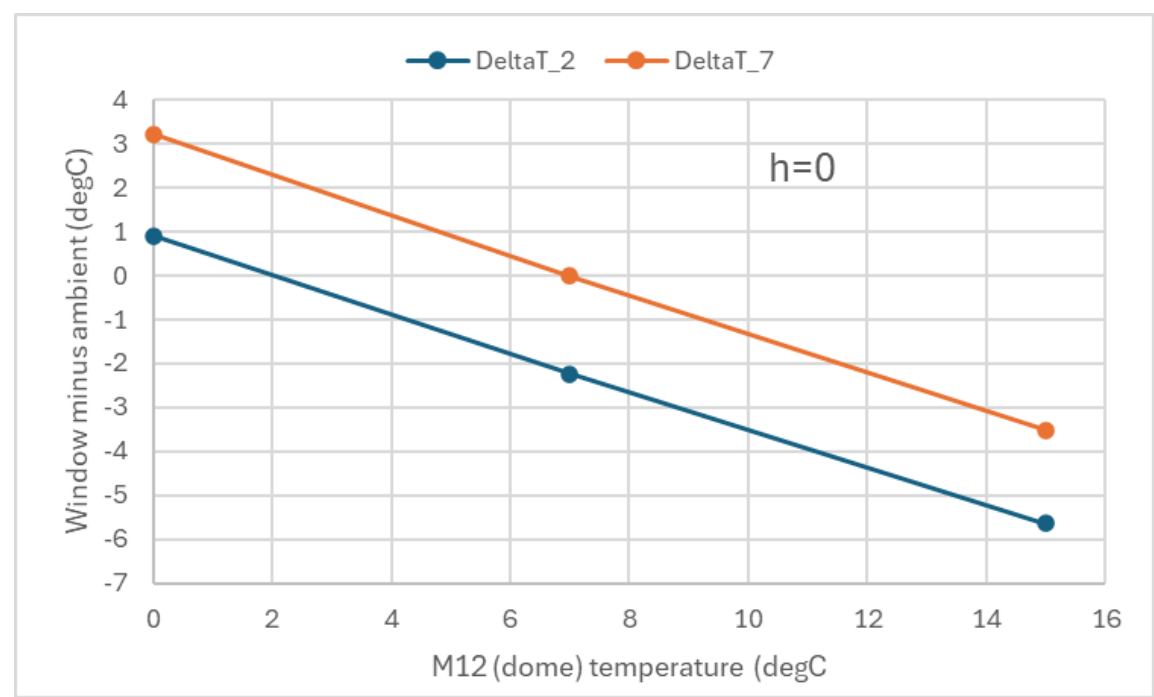

Figure 9. Differential between window temperature and ambient temperature as function of external temperature for internal temperatures of 2 and 7 degC.

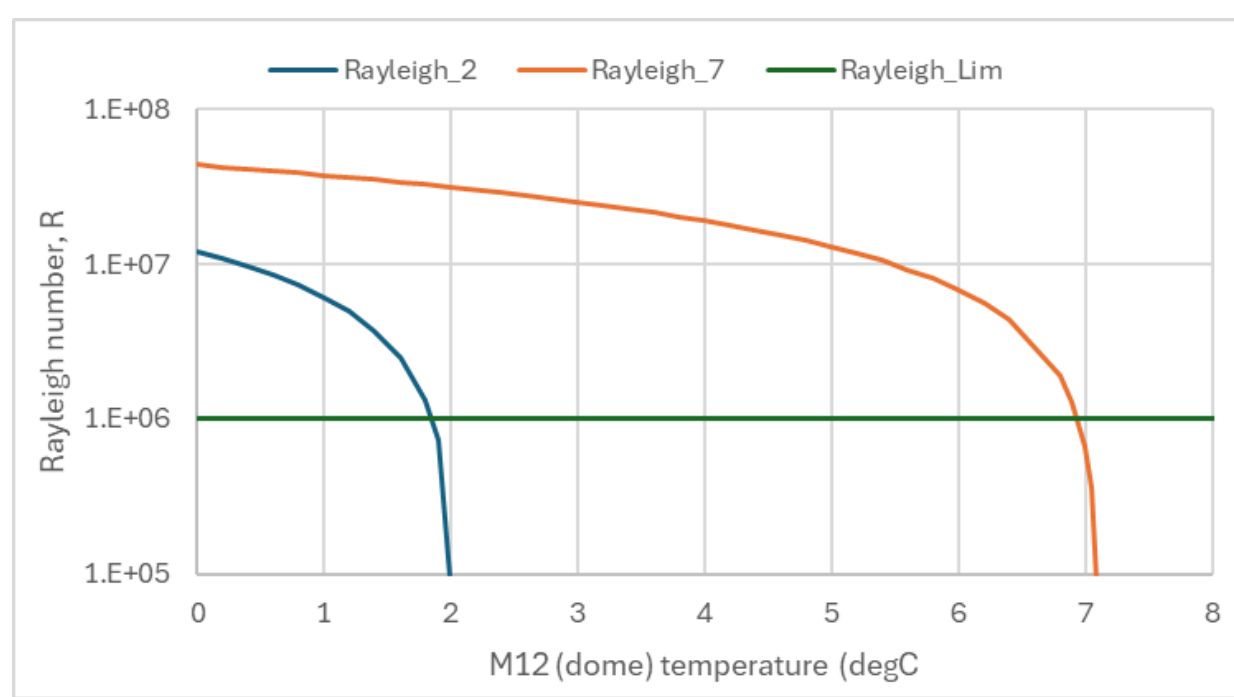

Figure 10. Estimated Rayleigh number for the space between the NGSS window and M12 as function of dome temperature.

Internal turbulence should be avoided during operation since it will impact observations. The space between NGSS window and M12 corresponds to very high-altitude layers (well outside the earth's atmosphere) and it is not seen by the tomographic wavefront sensors operating on the laser guide stars. Condensation should also be avoided since it also affects observations, nut also might affect the quality of optical coatings and since it is likely to leave traces on the window surface due to dust coagulation.

To mitigate these effects, we will maintain the internal temperature at 2 degC, and implement a system connected to the filtered air supply for the NGSS internal flushing, to allow flushing of the window top surface in case of excessive humidity.

### 3.3 Taking on-board control electronics

Although not part of the rescope exercise as such, reshaping HARMONI was also the opportunity to make an important change to the product breakdown structure (PBS) of the instrument by handing electronics control over to the subsystems. This negated a fundamental choice made at the beginning of the project whereby all the electronics controlling motors and other devices were collected together into a separate system, the HARMONI control system, HCS. While this was seen as an advantage at the time, allowing electronics development to be concentrated in the hands of one of the consortium members specialized in control electronics, this situation, confusing the PBS with the work breakdown structure (WBS), turned out to be difficult to handle in terms of interface control. Indeed, every pinout in every connector had to be considered an external interface for the subsystems, adding large overheads in terms of documentation and discussions.

In practical terms, however, individual control electronics specialists were dedicated to specific subsystems and essentially formed part of the subsystem teams, building up the control electronics in direct collaboration with the teams. The design actually evolved thanks to interactions and shared documents rather than based on configuration-controlled interface drawings. The new PBS, see Figure 11, therefore officialised the way things already functioned. Of course, in terms of management of individual systems and subsystems, this was still a quite significant change since it handed design and build responsibility over to the system and subsystem managements and forcing certain partners to look for new resources.

For NGSS, the rescope was also the opportunity to reconsider the volume available for electronics. Space constraints in the previous version limited the cabinet capacity to six small cabinets tucked in between the NGSS structure and the dome of the cryostat. Huge efforts were made by the electronics designers to fit all the electronics, but only at the cost of certain shortcuts, like mutualized cabinet control and zero margin. It was found that going to the new HARMONI, it was possible to widen the NGSS swept volume envelope sufficiently to hang reasonably-sized cabinets outside of the NGSS structure allowing a doubling of the cabinet height.

In the current NGSS, each subsystem has one or even two cabinets of which they are the owner and responsible for internal design and overall performance. They also own the harness and connector panels connecting control modules to each device. Their interfaces with the rest of the instrument are therefore reduced to power, glycol, ethernet and Ethercat connections, see Figure 12.

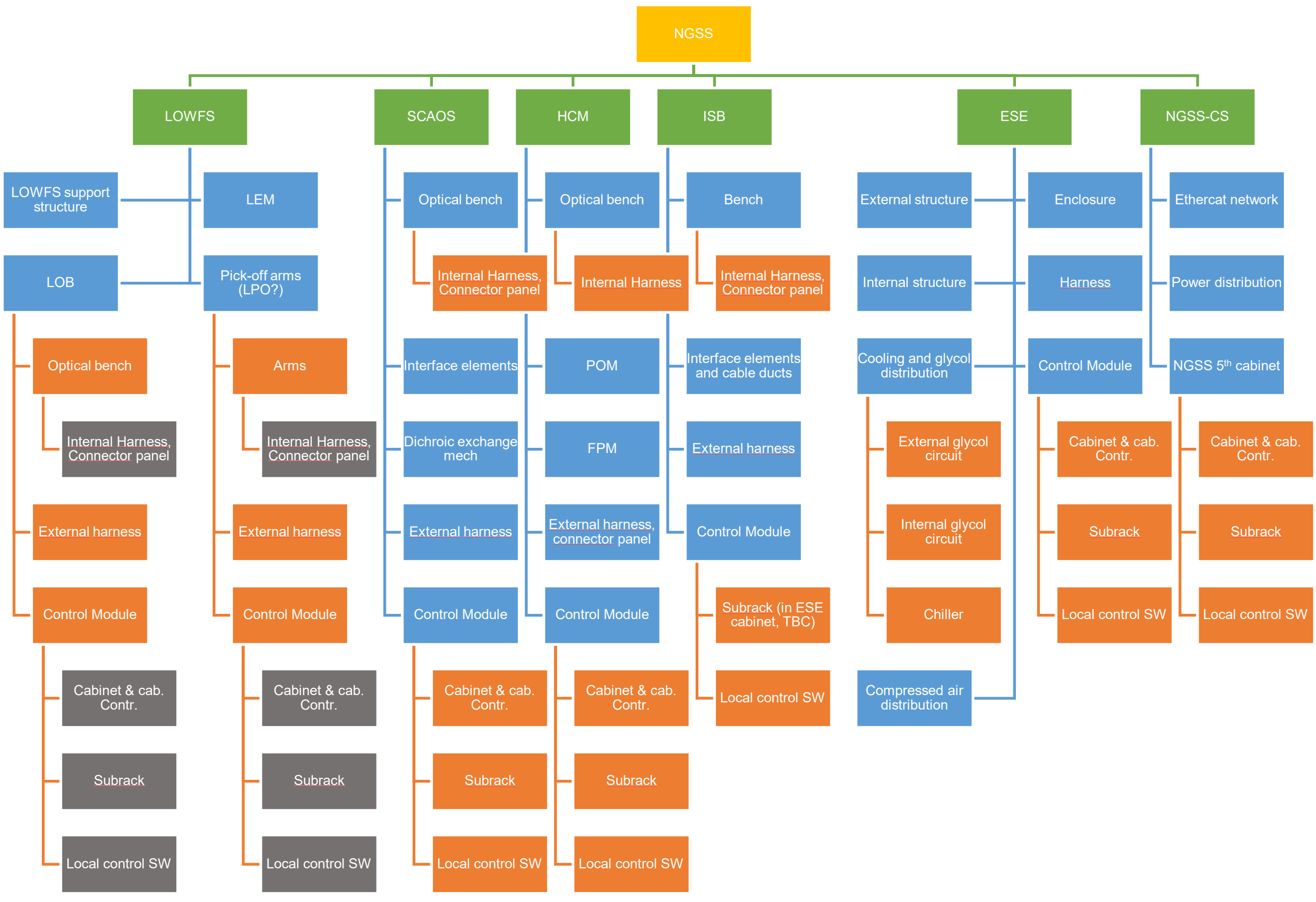

Figure 11. NGSS product breakdown structure (PBS) focussing on the new control electronics elements.

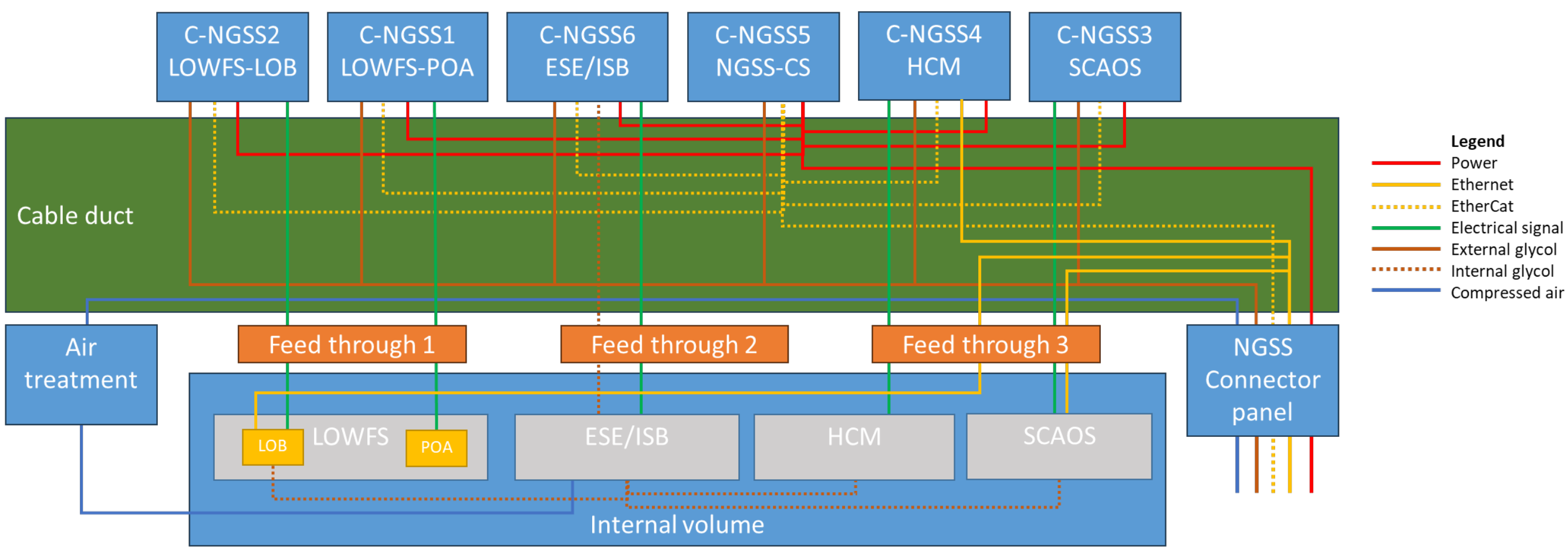

Figure 12. NGSS connectivity diagram.

## 4. RESCOPED SUBSYSTEMS

The effects of rescope on each one of the NGSS subsystems are outlined in the following sections, outlining the differences with respect to the previous NGSS paper [4]. Further references to individual subsystem reports are indicated when these are available.

### 4.1 ESE and ISB: External and internal supports and enclosure

Going to the second MORFEO output port, as described above, implied closing off the cooled and protected NGSS volume by introducing an entrance window since the large optical relay, the FPRS, with which we earlier shared the volume, no longer exists. To properly enclose all the NGSS internal organs, including the SCAOS field mirror which entered into the FPRS volume[4], the top cover had to be lifted up by some 400 mm in the central region. However, to stay clear of the beam emanating from the MORFEO output port, the cover had to be lowered by some 100 mm around the rim. To achieve this, a conical or “Chinese hat” shape was introduced, whose practical implementation imposed a dodecagonal architecture, see Figure 13. Together with the presence of larger cabinets, now fixed to the outer structure of the ESE subsystem rather than tucked in between the NGSS volume and the cryostat dome, this new design of ESE changes quite significantly the external aspect of the NGSS system.

The internal support bench (ISB) remains essentially unchanged after rescope, maintaining its interfaces with ESE, to which it’s mounted on one side and the subsystems which are mounted onto it, SCAOS and HCM, on the other.

### 4.2 LOWFS: Low-order WFS for multi-conjugate adaptive optics, MCAO

LOWFS has gone through wide-ranging changes over the years, as the number of guide probes has gradually increased from one to three, following increased understanding of telescope and instrument limitations and now the needs for MCAO replacing LTAO. The shoulder-elbow architecture of the guide-probe design is maintained from pre-rescope, for which a prototype was already well advanced, see Figure 14. The prototype is nearing completion and intermediate results are available, see [9].

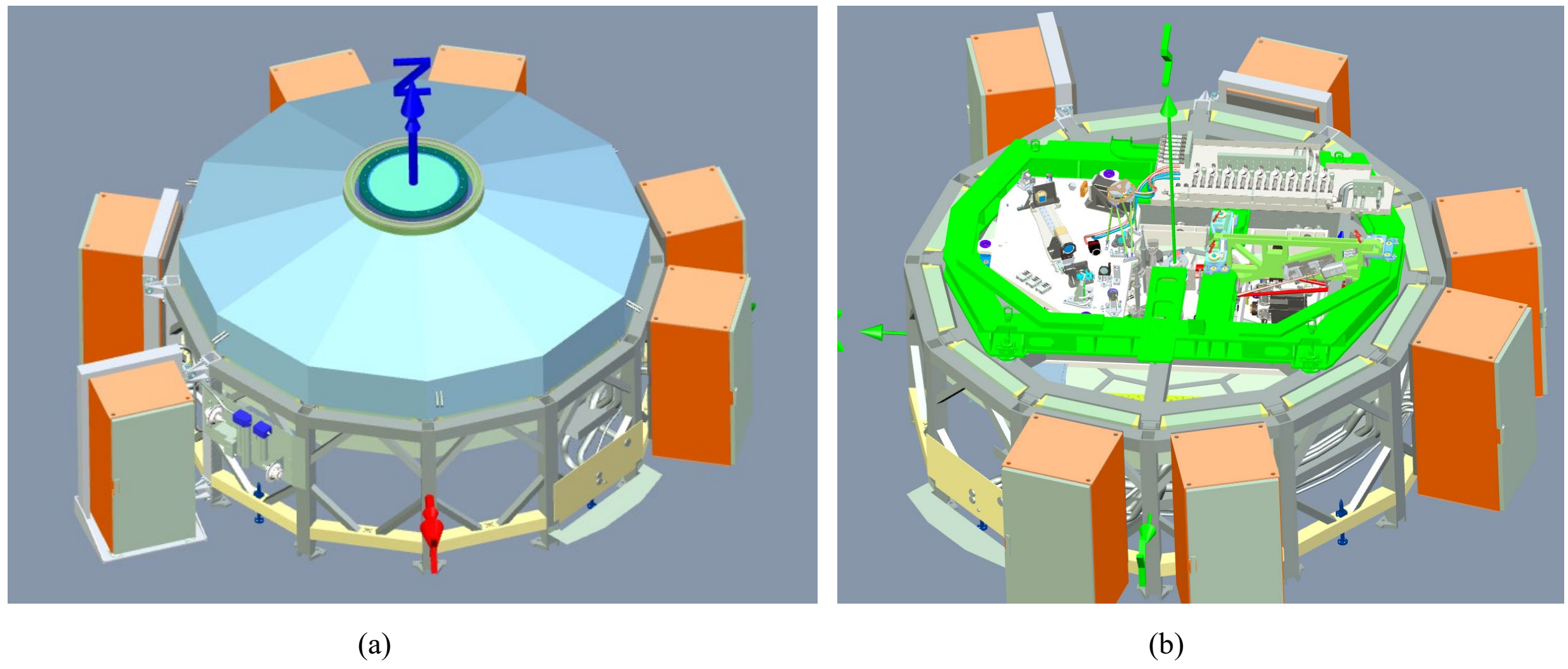

(a) (b)

Figure 13. Rescoped NGSS, showing the new “Chinese hat” design of the ESE and the new, much larger cabinets (a). Removing the top cover (b), the ISB (highlighted in green) carries the SCAOS (left) and HCM (right) subsystems as well as the manifold for internal cooling distribution.

LOWFS has two types of sensors in each arm, whose optical paths are laid out in the optical bench, see Figure 15:

- Truth sensor (TruthS), an 8x8 Shack-Hartmann sensor working in the 0.6 to 1um band using an ALICE camera
- Fast tip-tilt and focus sensor (TTFS) working in the J and H bands (1.2 to 1.8um) using a FREDA camera

The TTFS sensors are equipped with an ADC and a focusing stage for optimal diffraction limited imaging of the reference object.

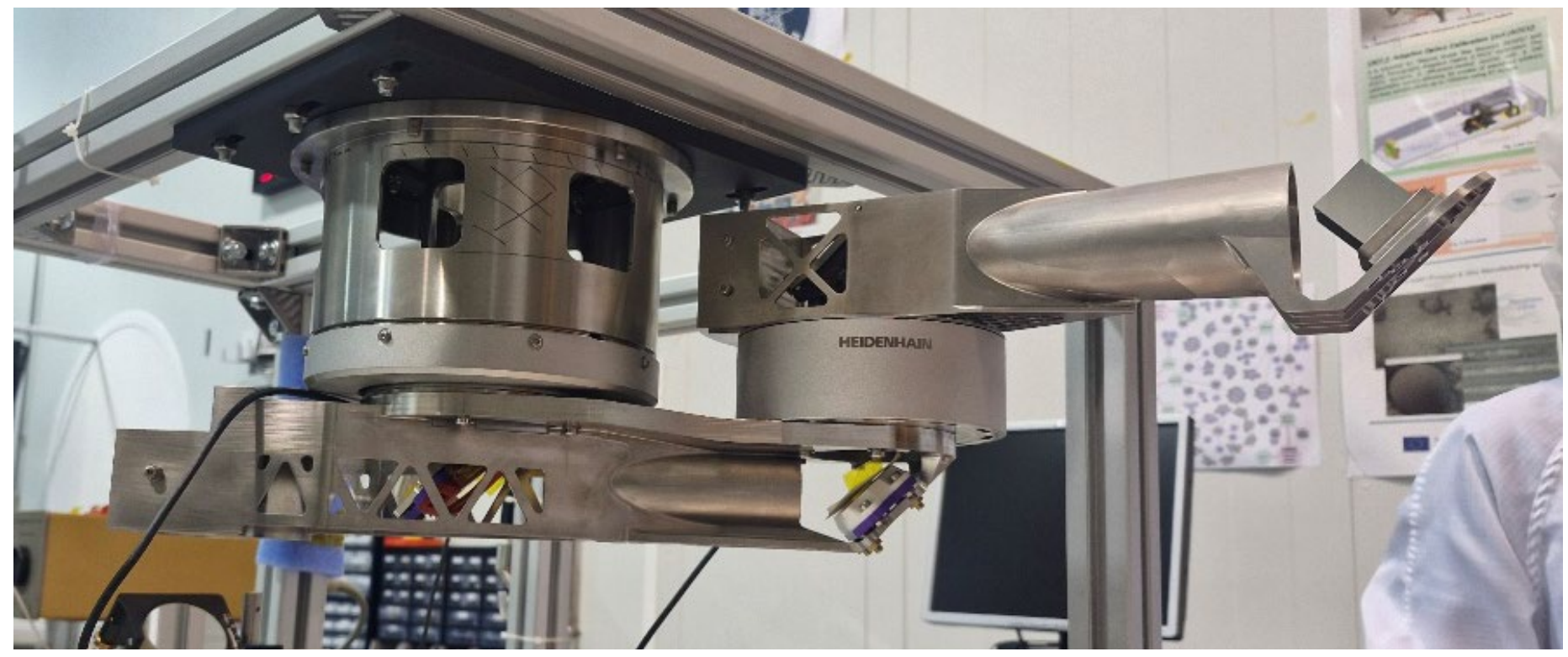

Figure 14. The LOWFS pick-off arm prototype.

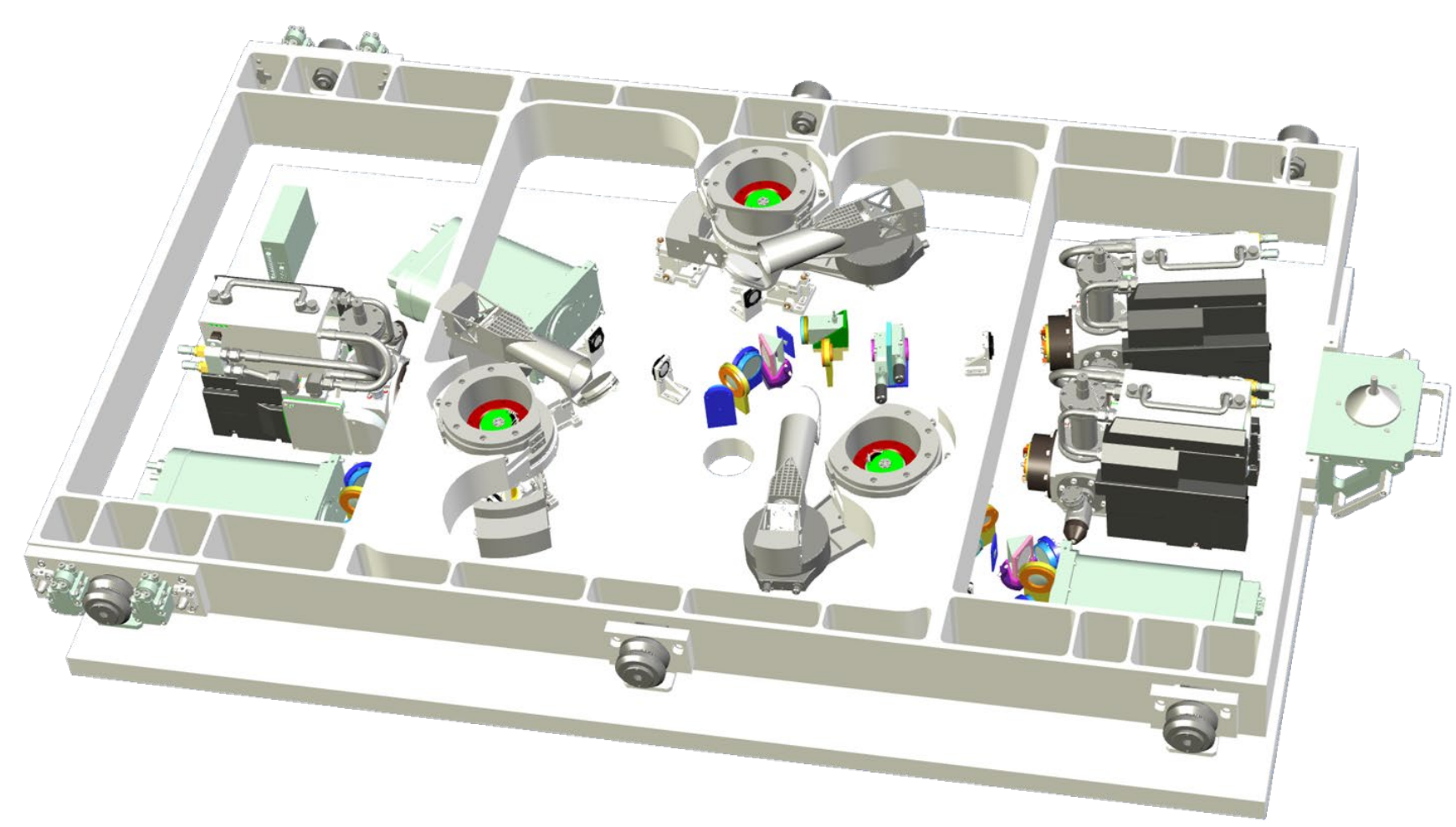

Figure 15. The LOWFS opto-mechanical bench including 3x FREDA cameras for fast Tip-Tilt Focus sensors and 3x Alice cameras for Truth sensing.

### 4.3 SCAOS: Sensors for single conjugate adaptive optics

SCAOS has survived the rescope with only minor changes [10]. The main changes to its requirements are a reduction of the patrol field from 30’’ to 10’’ diameter and an increase in the shortest observing wavelength from 850 nm to 1035 nm. This allowed a reduction in size and complexity, hence manufacturing cost and risk, of the main dichroic plate and the freeform field mirror, and the removal of the short-wave dichroic which had allowed observations down to 850 nm with a reduced patrol field. SCAOS is therefore left with only two dichroic plates on its exchange mechanism, the main dichroic for SCAO observations and the smaller high-contrast dichroic, dedicated to HCAO observations. This latter is equipped with an assembly of mirrors and lenses allowing transmission of the science beam towards the high-contrast module (HCM) and to reinject it back into the beam towards the IFS. The reduced patrol field also reduces the travel range for the object selection mirror, making this component less critical in terms of the stringent world coordinate system (WCS) requirements for the instrument.

The team also chose to use the ALICE camera, now deemed sufficiently mature, instead of the OCAM2k for the pyramid wavefront sensor.

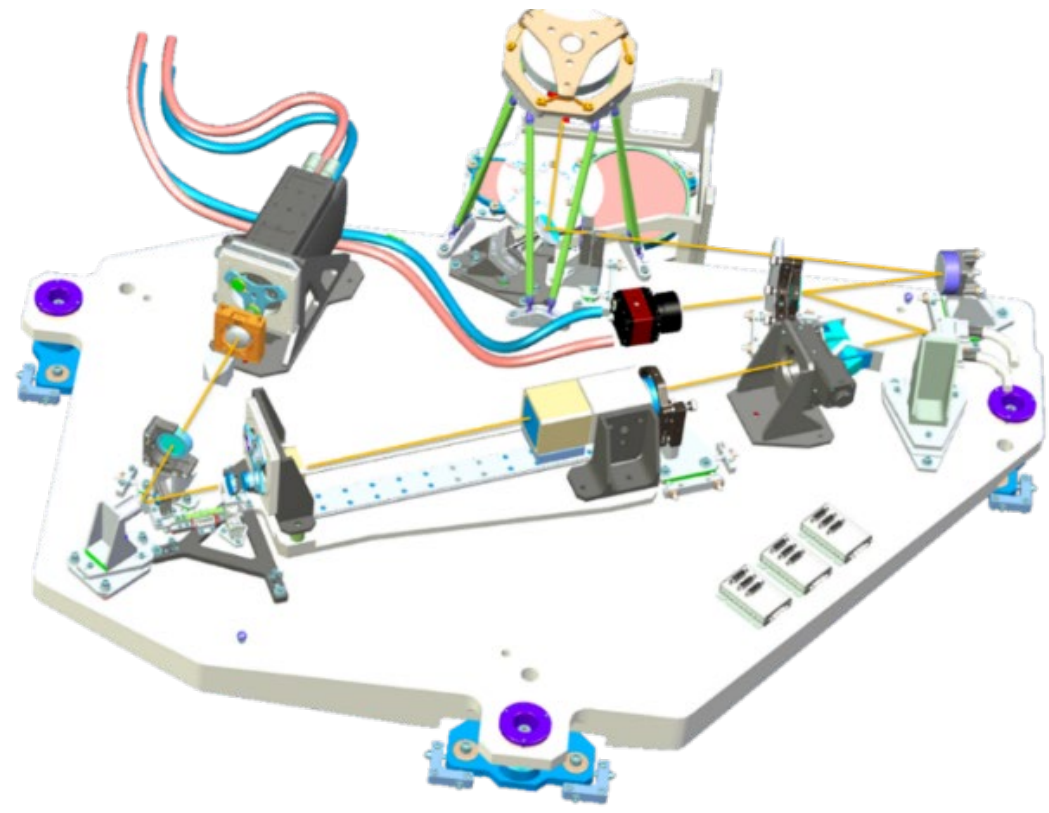

Figure 16. The SCAOS optical bench.

### 4.4 HCM: High-Contrast Module

Following rescope, the HARMONI High-Contrast imaging mode (HCAO) has been upgraded from a goal to a required functionality. While this ensures its presence in the final instrument, it does not impact the design of the subsystem that implements this functionality, the HCM [11]. However, rescope also modified the final image plate scale of the science focal plane, increasing the smallest scale from 4 mas/spaxel to 6 mas/spaxel. Since high-contrast imaging requires Nyquist sampling at the shortest observing wavelength, leading to a need for a 3.88 mas/spaxel scale, the corresponding image magnification needs to be performed in the HCM optical train. A small magnification factor was already designed in before the rescope, but this has now been increased by some 30%. The optical and mechanical redesign work is ongoing, but the resulting design remains very similar to the original one, see Figure 17.

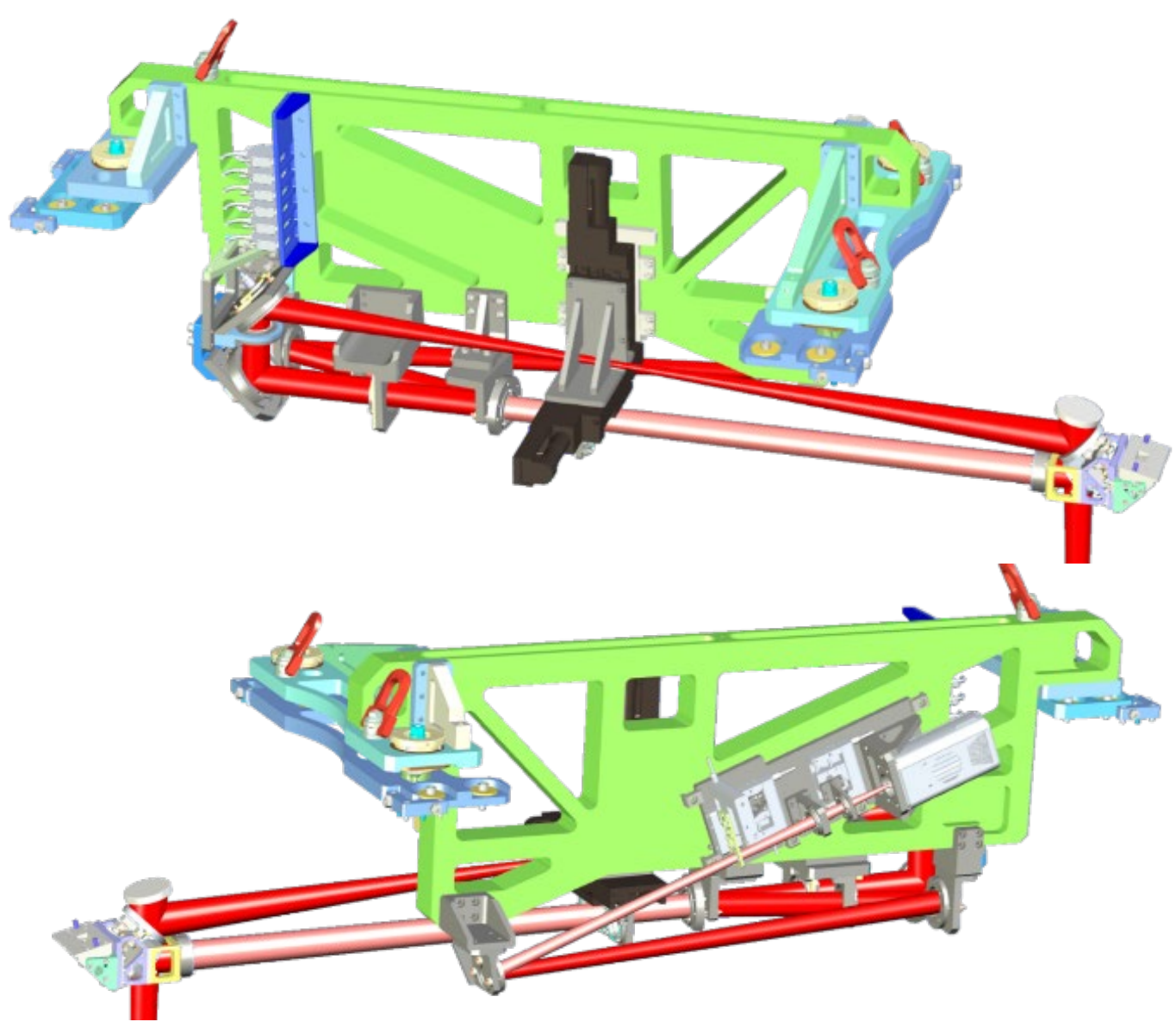

Figure 17. Bi-lateral views of the high-contrast module (HCM) showing the science path (top) and Zelda wavefront sensor path (bottom).

### 4.5 NGSS-CS: Control system

In parallel with the rescope, an important change in the instrument PBS was decided. All local control electronics, housed in six cabinets mounted onto the circumference of NGSS, was handed over from the centralized HARMONI control system (HCS) to the systems and subsystems. This greatly simplifies the external interfaces, now limited to power, Ethercat and Ethernet, in addition to software interfaces, rather than also including detailed, pin-by-pin, electrical control interfaces for each device. Each subsystem now has its own cabinets for which they are fully responsible in terms of internal architecture and performance verification.

Common electronics and control, including power and network distribution, is handled by a new subsystem, the NGSS-CS. The work package of this subsystem also includes the design of a specific cabinet model for NGSS, fitting the available volume and providing the required thermal insulation and control elements. Once this design is confirmed, it will be built for that subsystem and offered as a catalog item for each of the other subsystems to order.

All electronics design is implemented within the Eplan environment [12], allowing outputs in terms of schematics, see Figure 18, and 3D drawings of cabinet implementations, see Figure 19.

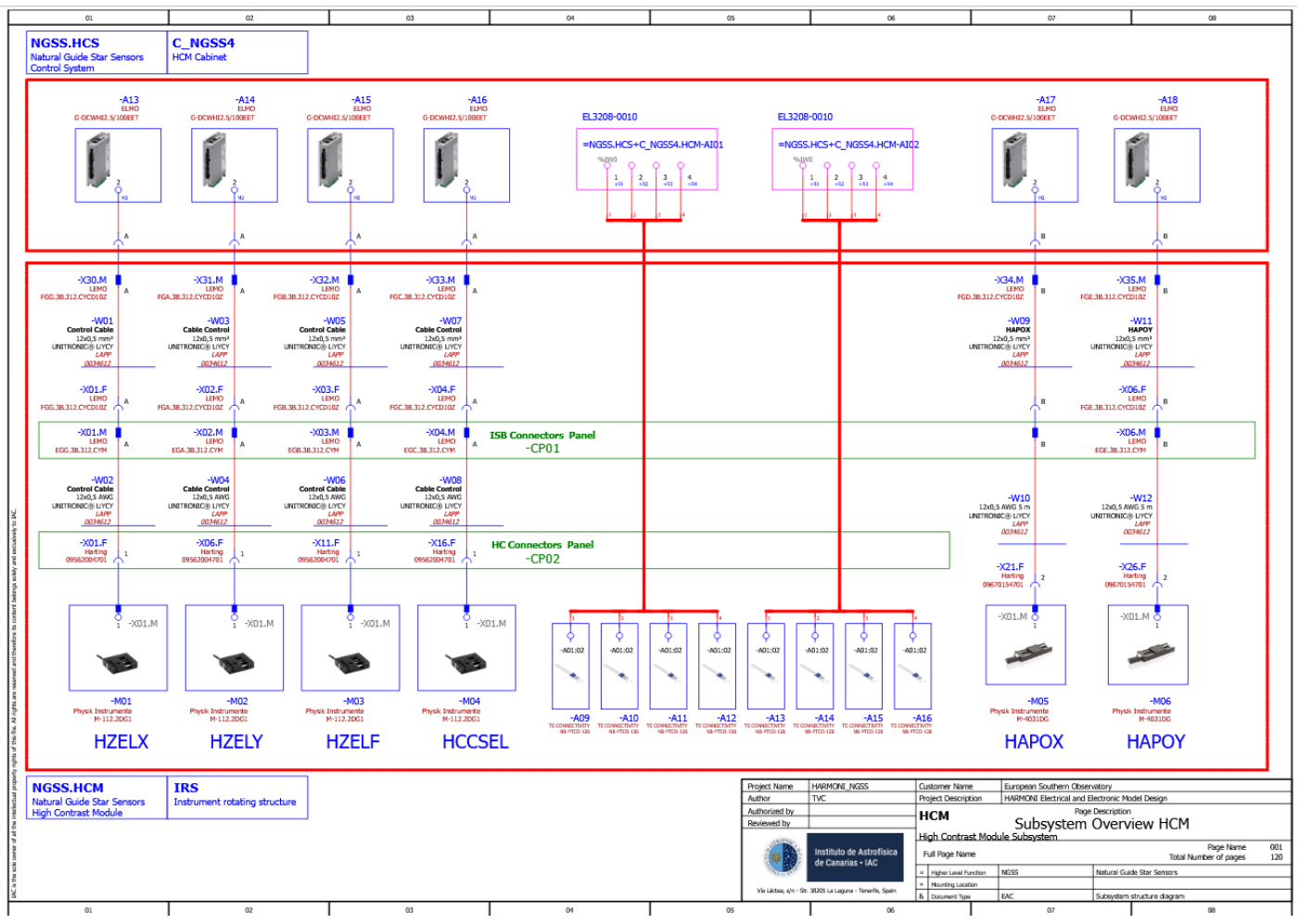

Figure 18. Typical electronic diagram output from ePlan.

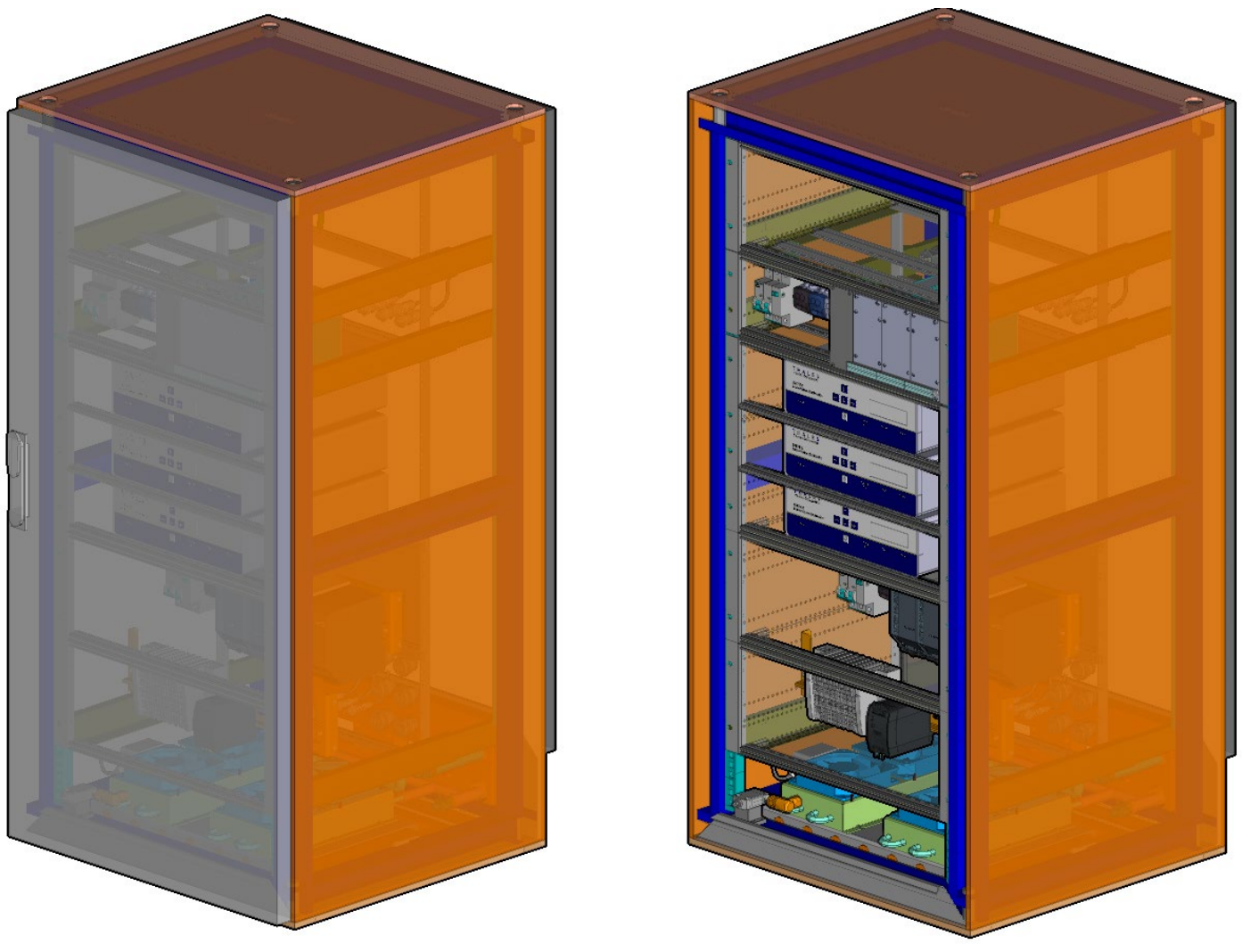

Figure 19. Preliminary implementation of one of the NGSS cabinets with (left) and without (right) the front door.

## 5. CONCLUSIONS

Having finished the process of rescoping HARMONI, this paper describes the changes implemented in the natural guide star (NGSS) system. The most important changes are seen in the external support and enclosure (ESE) subsystem due to changes in the global instrument design, and in the low-order wavefront sensor (LOWFS) subsystem due to the change from laser tomographic AO (LTAO) to multi-conjugate AO (MCAO). Another important change was the increase in plate scale, leading to a modification of the high-contrast module (HCM) to ensure Shannon sampling. Also, the high-contrast observing mode (HCAO) was promoted from a goal to a requirement, but this has no major impact on the design since all its necessary features were already designed into the subsystem. Finally, in parallel with the rescope process, a change in product breakdown structure shifted all local control hardware and software from the centralized HARMONI control system (HCM) over to individual systems and subsystems. The NGSS subsystems are therefore now responsible for their own cabinets, and a new subsystem, the NGSS-CS has been created.

Dust settling after the upheaval caused by the rescope, the NGSS is now firmly back on track towards FDR. It has been painful, but the instrument has clearly gained in terms of science focus and performance, and the teams are eager to continue the development of this great machine in view of magnificent new astrophysical discoveries.

## ACKNOWLEDGEMENTS

This work benefited from the support of the French National Research Agency (ANR) under the Future Investment Program F-CELT (ANR-21-ESRE-0008) and the Commission Spécialisée Astronomie-Astrophysique (CSAA) of the INSU-CNRS.